\documentclass[conference]{IEEEtran}
\usepackage{graphicx}
\usepackage{subfig}
\usepackage{epstopdf}
\usepackage[linesnumbered,lined,boxed,ruled ]{algorithm2e}
\usepackage{multirow}
\usepackage{amsmath}
\usepackage{array}
\usepackage{microtype}
\usepackage{balance}
\usepackage{color}
\usepackage{cite}
\usepackage{amsmath,amssymb,amsfonts}
\usepackage{capt-of}
\usepackage{algorithmic}
\usepackage{textcomp}
\usepackage{xcolor}
\def\BibTeX{{\rm B\kern-.05em{\sc i\kern-.025em b}\kern-.08em
    T\kern-.1667em\lower.7ex\hbox{E}\kern-.125emX}}
\begin{document}

\title{Multibit Tries Packet Classification with Deep Reinforcement Learning}

\author{\IEEEauthorblockN{ Hasibul Jamil, Ning Weng}
\IEEEauthorblockA{\textit{Electrical and Computer Engineering Department} \\
\textit{Southern Illinois University, Carbondale, USA }\\
Email: \{mdhasibul.jamil,nweng\}@siu.edu}
}

\maketitle
\thispagestyle{empty}
\begin{abstract}

High performance packet classification is a key component to support scalable network applications like firewalls, intrusion detection, and differentiated services. With ever increasing in the line-rate in core networks, it becomes a great challenge to design a scalable and high performance packet classification solution using hand-tuned heuristics approaches. In this paper, we present a scalable learning-based packet classification engine and its performance evaluation. By exploiting the sparsity of ruleset, our algorithm uses a few effective bits (EBs) to extract a large number of candidate rules with just a few of memory access. These effective bits are learned with deep reinforcement learning and they are used to create a bitmap to filter out the majority of rules which do not need to be full-matched to improve the online system performance. Moreover, our EBs learning-based selection method is independent of the ruleset, which can be applied to varying rulesets. Our multibit tries classification engine outperforms lookup time both in worst and average case by 55\% and reduce memory footprint, compared to traditional decision tree without EBs.

\end{abstract}


\begin{IEEEkeywordsname}
	packet classification, machine learning, optimization
\end{IEEEkeywordsname}


\section{Introduction}


Packet classification is a key function to support network applications like firewalls, intrusion detection, and differentiated services and OpenFlow switch. The problem of packet classification is similar to point location problem in a multi-dimensional geometric space.  The meta data in a packet, (i.e., packet headers) contains different fields, representing different dimensions in space. Given a packet, finding out where exactly that packet is located in that space is essentially the packet classification problem. A classifier is a set of rules, each rule specifies a pattern (i.e. values or range of values) on different fields of a packet header. In this way, all these rules could be represented as hypercubes in that same space.

Existing algorithmic solutions to packet classification include decision-tree-based techniques~\cite{gupta:99hicuts} and decomposition-based techniques~\cite{Lakshman98high-speedpolicy-based}\cite{Baboescu01_ABV}. Meanwhile, several orthogonal solutions for packet classification have explored, including \cite{Lossy_15} using the prefix probability, \cite{Jatrie_14} using entropy for a compact data structure, \cite{Rott_Compressing_InfoCom13} using compressing tables.

However most of these solutions are built on heuristics (e.g., increasing split entropy~\cite{gupta:99hicuts}, balancing splits with custom space measures~\cite{gupta:99hicuts}, special handling for wildcard rules~\cite{singh2003packet}) that fails to generalize the process of building a decision tree for different set of rules. On the other hand if these solutions are specifically tuned to exploit certain characteristics present in a given ruleset, those characteristics may not be present in another ruleset. As a result, this environment (i.e., ruleset) specific heuristics typically suffers with sub-optimal performance. Another drawback of this hand-tuned heuristics is the absent of a global objective (e.g., tree depth or the number of nodes in the tree). Their decision making is often based on local information (difference between the number of rules in the current node~\cite{Hsieh-manyfield-tnsm-19}, the number of different ranges in different dimensions ~\cite{singh2003packet} ). This local information is loosely related to the global objectives and that leads to their performance to be sub-optimal.

To address above mentioned limits of heuristics based solutions, we will build a decision tree using deep learning approach. The promising aspect of deep learning in systems and networking problems~\cite{MoDong2018, HongziMao2017}, inspires us to use deep learning in packet classification. Essentially we aim to use learning-based approach to generate an high performance and low memory packet classification engine for any ruleset without relying on heuristics.

In this paper, we present multibit-tries packet classification engine, which is generated by deep reinforcement learning. For a given set of rules, our solution employs a learning-based approach to find out the effective bits. Effective bits (EBs) are essentially the selected bit positions from a 5 field, 104-bit tuple structure, that divides the original rulesets into multiple groups. We use  these EBs to retrieve a large number of possible candidate rules in just one memory access. Therefore, deep reinforcement model generates an optimized decision tree for a given ruleset and using that decision tree we calculate required effective bits. The selected effective bits are then used to traverse the original decision tree in multibit tries approach that yields up to 55\% performance improvement for different set of ClassBench~\cite{Taylor_ClassBench_2007} rules.

In summary, there are three main contributions of this work:

\begin{itemize}
  \item A  deep learning based multi effective bit selection method is proposed by leveraging the statistical characteristics in a ruleset to conduct multi-dimensional lookups.

  \item Based on effective bit, a multibit tries packet classification engine is designed for the system scalability in both processing throughput and storage requirement

  \item Our multibit-tire packet classification can achieve classification time improvement up to 55\% compared to unibit decision tree with small memory footprint improvement for varying ruleset.

\end{itemize}

The remainder of this work is organized as follows. Section \ref{sec:back_motive} gives background for related research on packet classification. Section \ref{sec:method} presents the proposed method. Experimental setup and results are shown in Section \ref{sec:results}. Finally, Section \ref{sec:conc} summarizes and conclude the paper.

\section{Background and Related Work} \label{sec:back_motive}

Packet classification is an important and hard problem~\cite{Taylor:ACMCS05}. Existing solutions can be classified into three categories: hardware solution, hand-turned heuristic algorithmic solutions and machine-learning based solutions. In this section, we will first review several heuristic algorithmic solutions, finally a brief review of machine learning based solutions.

Algorithmic solutions for packet classification include decomposition and decision tree. Decomposition-based solutions work on each field in a ruleset independently using cross-products~\cite{Taylo05_DCFL}\cite{Srinivasan98_crossproducting} or header chucks~\cite{Gupta:1999}\cite{Lakshman98high-speedpolicy-based} for the intermediate results. These solutions merge the results from different fields to produce the final match results. 
As discussed in~\cite{Gupta:2001}, the time complexity is $O(dW)$ for cross-product solutions and is $O(d)$ for header chuck solutions. The storage complexity is $O(N^d)$ for~\cite{Taylo05_DCFL}\cite{Srinivasan98_crossproducting}\cite{Gupta:2001} and $O(dN^2)$ for~\cite{Lakshman98high-speedpolicy-based}. Decision-tree-based approaches~\cite{Kennedy_LowP_ToVLSI14}\cite{Lim_BC_2014}\cite{Yang_D2BS_TOC14} analyze all fields in a ruleset to construct tree data structures for an efficient packet header lookup. Tree depth and rule duplication in a decision tree affect the searching efficiency and memory requirement of the implementation. For the matching process, decision-tree-based solutions traverse the tree using field values to make branching decisions at each node until a leaf is reached. 
According to the discussion in~\cite{Gupta:2001}, the growth of field numbers in a ruleset results in a linear increase of processing latency and the time complexity is $O(d)$. Based on the nature of decision trees, the rule duplication is carried over to the next layer. The growth of field numbers in a ruleset results in an exponential increase of memory requirement and the storage is $O(N^d)$.


Learning based approach could be divided into two categories. One learning-based approach could be getting rid of the decision tree itself, a neural network will output the matching rule for a packet, given the packet's header fields. It has shown that a deep neural network (DNN)  could be used to replace B-trees for indexing ~\cite{
}. This approach has some serious drawbacks. First, it doesn't assure 100\% accuracy which is an absolute requirement for most crucial packet classification services (e.g. firewalls, access controls). The reason for not having a 100\% accuracy is due to the fact that training a neural network is essentially a stochastic process. If a DNN replaces the tree, we still need another system to verify if the DNN result is correct or not. Secondly, a packet could match multiple rules in a ruleset so, given a packet, after the first match, system still  needs to go through the other rules to see if the first matching rule has the highest priority among all the matching rules or not. In addition, for a large ruleset, the required DNN models will be very large in size and very difficult to train to obtain high accuracy. The other category is rather unorthodox, involves building decision trees utilizing Deep learning (RL) which  has been introduced in ~\cite{Liang-NPC-2019}. There, authors shows how deep reinfocement learning could be used to generate optimized decision tree for any given ruleset. 
Reinforcement Learning can learn the most efficient heuristic for a given environment. 

Our proposed solution falls into learning based category and we introduces an effective bit selection scheme from a decision tree and using those effective bits in mutibit tries we improve the classification performance and memory footprint per rule.

\section{Methodology} \label{sec:method}


An learning based system is required to tackle the problem of generating optimal decision trees for different given environment (i.e., rulesets). Once the optimal decision tree is built, we introduce a multibit lookup scheme that generates significantly lower memory footprint and classification time. This multibit lookup scheme finds some number of bit positions in the packet header (i.e., ith \& jth bit of SrcIP and/or mth \& nth bit of DstIP) to generate a bitmap and essentially enables to retrieve a large number of classifier rules in just one memory access. We called these bits as effective bits (EB). For a decision tree, these processes stops when all decision tree leaves have no more than $binth$ (bin threshold) rules, $binth$ controls the amount of linear searching at the end of the tree search.

For a decision tree, we reduce the size of the tree by truncating a non-leaf node only if its total number of rules exceeded above a given group-threshold. We named this threshold as group-threshold because it groups a number of internal and leaf nodes together. 

In our scheme all the found effective bits are concatenated to form index of a lookup table. This lookup table is pre-computed and stored in memory, so for an incoming packet the value from the effective bit positions is calculated and use that value as index to search in the lookup table. Doing this, our scheme filters out the majority of non-related rules and only conduct costly matching against highly related matching rules. 

\subsection{Packet Classification Engine Overview}

\begin{figure}[t]
	\centering
	\includegraphics[width=.95\linewidth]{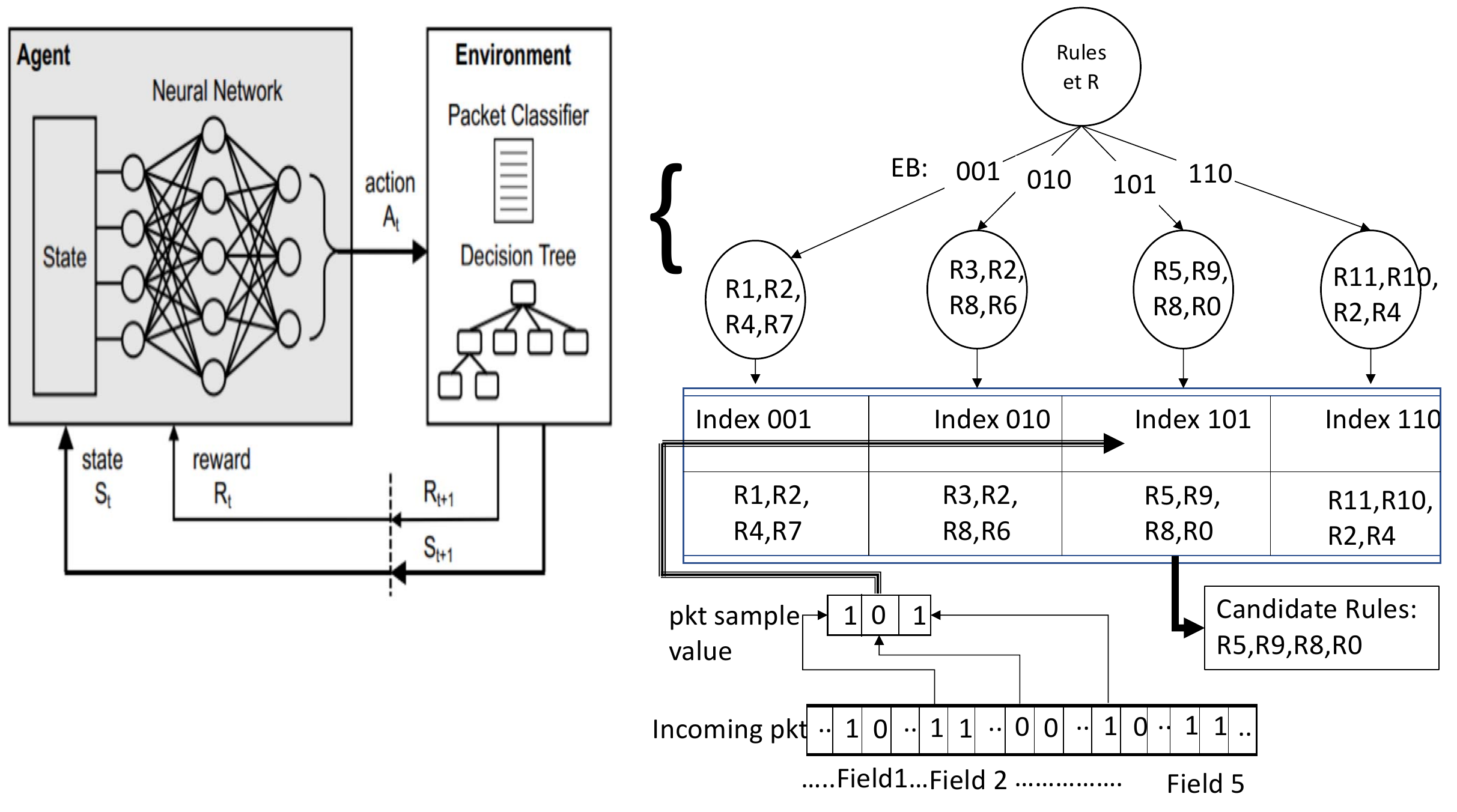}
	\caption{Packet classification engine.}
\vspace{-5mm}	\label{fig:overall-arch}

\end{figure}

As shown in Figure~\ref{fig:overall-arch}, a RL system consist of an agent and an environment, where agent repeatedly interact with the environment. The environment consists of a set of rules and a decision tree. Environment provides the current state $St \epsilon S$  which correspond to the current status of the decision tree. The agent receives this state information and uses a DNN model to choose an action $At \epsilon A$, i.e. cut or partition based on a policy. The state and action space are defined in the environment itself. A cut action divides a node along a chosen dimension (i.e., one of SrcIP, DstIP, SrcPort, DstPort, and Protocol) into a number of sub-ranges (i.e., 2, 4, 8, 16, or 32 ranges), and creates that many child nodes in the tree. A partition action divides the rules of a node into disjoint subsets (e.g., based on the coverage fraction of a dimension), and creates a new child node for each subset. Depending on the action taken by the agent, the environment also provides a reward signal $R_{t}$. Here, the goal of the model is to learn an optimized single policy  $\pi(a \mid s)$, where $a$ is the action and $s$ is the given state so that the cumulative reward after building the tree is maximized. These steps repeated for next time step $(t+1)$ and the tree build up incrementally. In summary, this decision tree building process could be casted as RL problem: the environments state is the current decision tree, an action is either cutting a node or partitioning a set of rules, and the reward is the classification time, memory footprint, or a combination of these two. Agent starts with an initial random policy, evaluates this policy with several roll-outs and then update the policy from the rewards of the roll-outs. A roll-out is a sequence of actions that builds a decision tree. All these actions are driven by a policy and a reward is received after completion of the tree. This process continues until the reward matches with the objective value. 


\subsection{Decision Tree via Deep Reinforcement Learning}

One interesting fact that could be leveraged on is that the action on a node is entirely depends on the node state itself not the state of the tree. If the sub-tree rooted at a node could be optimized, recursively the tree rooted from root node could be optimized (e.g., the memory access time and memory footprint of the tree could be optimized). The worst condition classification time is essentially the height of the tree considering the matching rule is in the farthest leaf node. And the memory footprint is directly related to the number of nodes in the tree.

\begin{figure}[t]
	\centering
	\includegraphics[width=.95\linewidth]{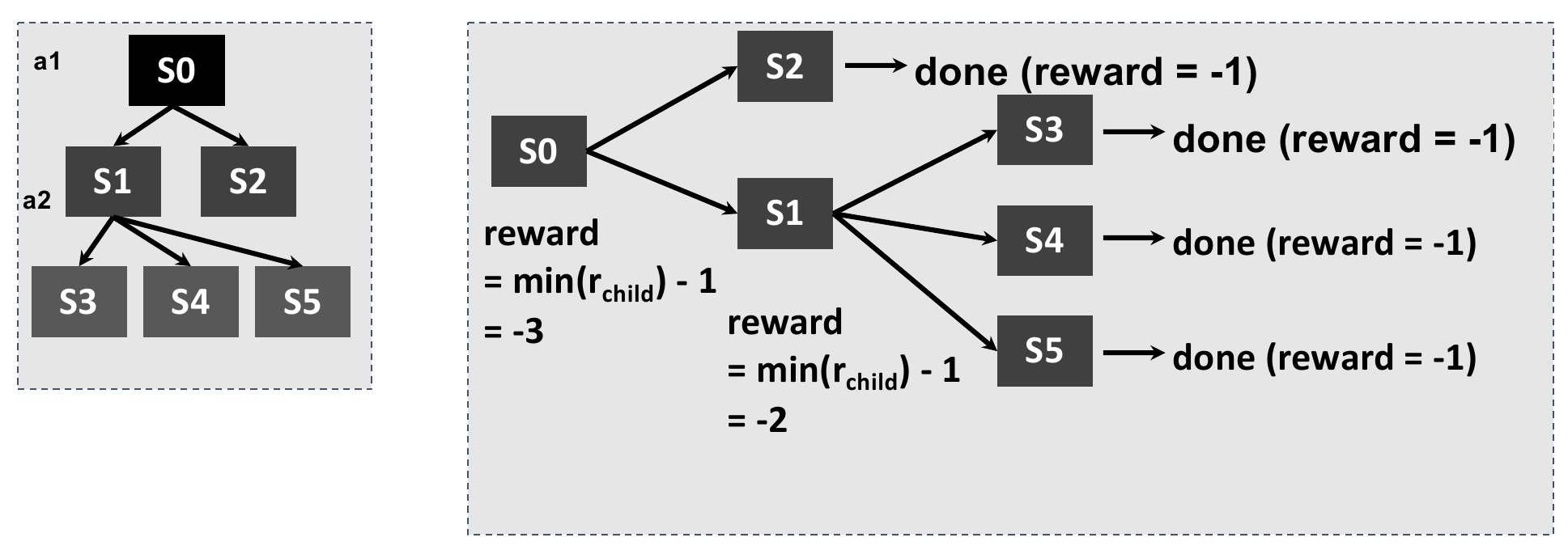}
	\caption{Illustration of reward of machine learning.}
	\label{fig:reward-cal}
\vspace{-5mm}
\end{figure}

The reward signal $(Rt)$ accommodate these two requirements for an action taken to optimize the global objective function of building a performance and memory optimized tree.
In this problem formulation, the environment is considered as a series of 1-step decision problems, each step yielding a reward. We call this secondary award and the actual or primary reward for these 1-step decisions is calculated upon completion of relevant sub-tree. Calculation of rewards are done not by summing over time but aggregating across tree branches. This is shown in Figure~\ref{fig:reward-cal}.

Considering a root node $S0$ in Figure~\ref{fig:reward-cal}, based on current policy, the agent decides to take action $a1$ to split $S0$ into $S1$, $S2$. Of these child nodes, only $S1$ needs to be further split (via a2), into $S3$, $S4$ and $S5$. $S2$, $S3$, $S4$ and $S5$ are leaf nodes. The experiences collected from this roll-out consist of two independent 1-step roll-outs: ($S0$, $a1$) and ($S1$, $a2$). The total reward R for each roll-out would be -3 and -2 respectively. It is important to mention that there is $O(log(n))$  delay between action and reward signal in this approach (where $n$ is the total number of nodes of the tree).

\subsection{Effective Bit Derivation}

\begin{figure}[h]
	\centering
	\includegraphics[width=.95\linewidth]{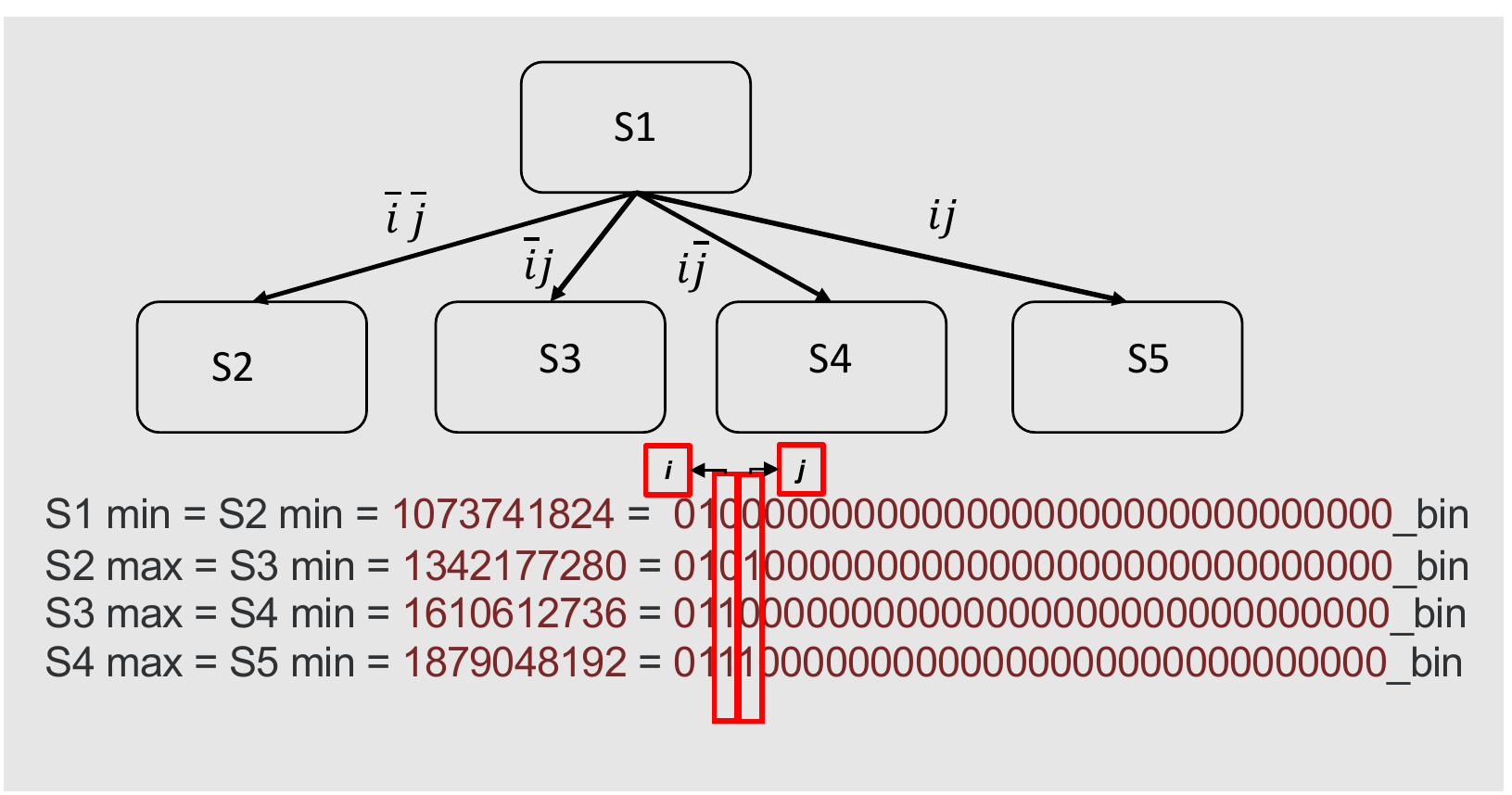}
	\caption{Illustration of effective bit selection.}
	\label{fig:effective-bit-example}
\vspace{-5mm}
\end{figure}

Once the optimized decision tree for a given ruleset is built, we could further improve its performance by incorporating an idea called concatenated EBs. The process of find the concatenated EBs is shown in Algorithm~\ref{alg:decisintree-EFbits} where a decision tree is given as the input. Each node of the tree has its attribute objects (i.e., number of child nodes, number of total rules, a special variable called $EBSetvalue$ storing bit positions pointing that node and its parent node.) All the  nodes pointers are arranged in such a way that a depth-first-search (DFS) could be performed (line 1-2). Every node of the tree is then traversed and checked for the selected attributes (line 3-8). Each node contains ranges of five tuples (i.e. min and max value of each dimension) covering all its rules. Whenever a cut action is done to a non-leaf node, the spawned nodes EBSetvalue variable is updates with corresponding bit positions (line 9). To illustrate this operations., let's consider an example. Given a node n, when we cut it to spawn other nodes, if n's state are {SrcIPMin, SrcIPMax, DstIPMin, DstIPMax,SrcPortMin, SrcPortMax, DstPortMin, DstPortMax, ProtocolMin, ProtocolMax} (e.g., all the rules n contain, SrcIPMin is the lowest IP among all the source IP in the ruleset and SrcIPMax is the maximum), spawning to child nodes could be represented by bit positions as shown in Figure~\ref{fig:effective-bit-example}.  Source IP is a 32 bit number and following example illustrate the idea of finding effective bits in more detail.
If a node $(S1)$ with following bound spawned to 4 new nodes $(S2,S3,S4,S5)$ and all the numbers are 32 bit in length.
\begin{center}
	$(S1) 1073741824 (SrcIPMin), 2147483648 (SrcIPMax)$\\
	$(S2) 1073741824 (SrcIPMin), 1342177280 (SrcIPMax)$\\
	$(S3) 1342177280 (SrcIPMin), 1610612736 (SrcIPMax)$\\
	$(S4) 1610612736 (SrcIPMin), 1879048192 (SrcIPMax)$\\
	$(S5) 1879048192 (SrcIPMin), 2147483648 (SrcIPMax)$\\
\end{center}

This action of spawning 4 different nodes could be represented by bit positions $i$ and $j$, where $\forall$ $i,j$ the following condition must be true $0$ $\leq$ $i,j $
$\leq$ $103$  as shown in Figure~\ref{fig:effective-bit-example}. So  $EBSetvalue$ variable of $S2,S3,S4$,and $S5$ will be updated with bit position $i,j$.

\begin{algorithm}
	\SetKwData{Left}{left}\SetKwData{This}{this}\SetKwData{Up}{up}
	\SetKwFunction{Union}{Union}\SetKwFunction{FindCompress}{FindCompress}
	\SetKwInOut{Input}{input}\SetKwInOut{Output}{output}
	\Input{decision tree}
	\Output{decision tree with every non-root node represented by a set of bit positions $s$,\
		where 0$\leq$ {every member of $s$} $\leq$ 103 or updated $EBSetvalue$ variable}
	\BlankLine
	
	\tcp {start from root node $S_0$ of the given decision tree}
	Fetch pointers of all the nodes in the tree ;\\
	Arrange the node pointers to do depth first search;\\
	\tcp {$EBSetvalue$ is a object for every nodes containing that nodes representing bit}
	\For{all the nodes $S_i$}{
		\If{$S_i$ is non-leaf}{
			Find number of children of $S_i$;\\
			Find dimensions to cut;\\
			Find range of each dimensions;\\
			Find bit positions required to represent current node's child nodes;\\
			Update each child nodes $EBSetvalue$ with the bit positions
		}
	}
	\caption{The pseudo code to extract each nodes representational bit positions}
	\label{alg:decisintree-EFbits}
\end{algorithm}\DecMargin{1em}

\begin{algorithm}
	\SetKwData{Left}{left}\SetKwData{This}{this}\SetKwData{Up}{up}
	\SetKwFunction{Union}{Union}\SetKwFunction{FindCompress}{FindCompress}
	\SetKwInOut{Input}{input}\SetKwInOut{Output}{output}
	\Input{decision tree, group-binth($th_G$)}
	\Output{truncated decision tree }
	\tcp {start from root node $S_0$ of the given decision tree}
	Fetch pointers of all the nodes in the tree ;\\
	Arrange the node pointers to do depth first search;\\
	
	\For{all the nodes $S_i$ except root node }{
		\If{$S_i$ is non-leaf}{
			\If{number of rules in $S_i$ $\geq$ $th_G$}{
				Replace pointer from $S_i$'s parent to $S_i$ with $S_i$'s parent to $S_i$'s childs\\
				Delete $S_i$
			}
		}
	}
	\caption{The pseudo code to truncate selected nodes}
	\label{alg:TruncatedDecisinTree}
\end{algorithm}\DecMargin{1em}

\subsection{Multibit Trie}

\begin{figure}[h]
 	\centering
 	\includegraphics[width=.80\linewidth]{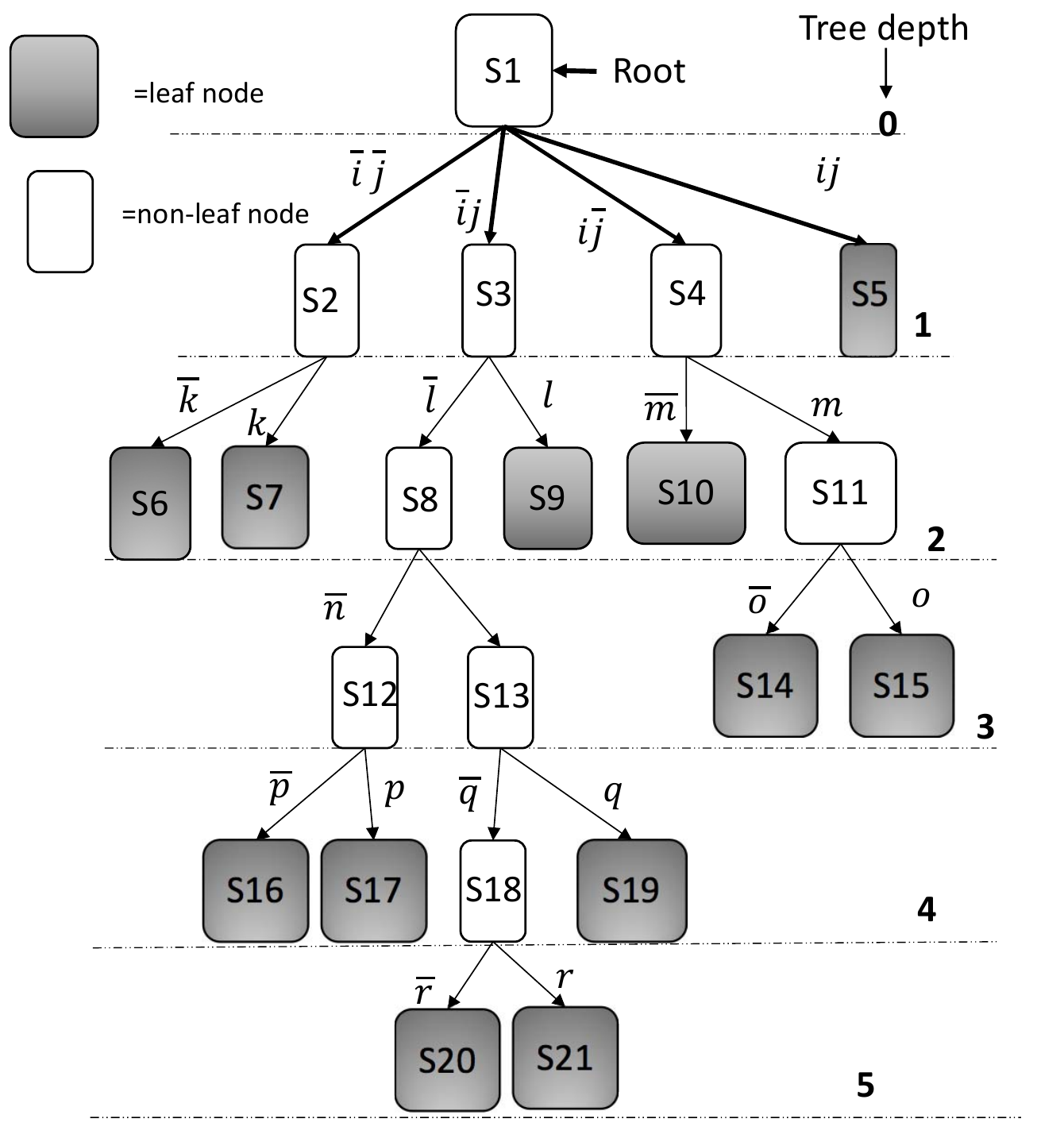}
 	\caption{Obtained Decision tree }
 	\label{fig:decision-tree}\vspace{-3mm}
 \end{figure}

\begin{figure}[h]
 	\centering
 	\includegraphics[width=.95\linewidth]{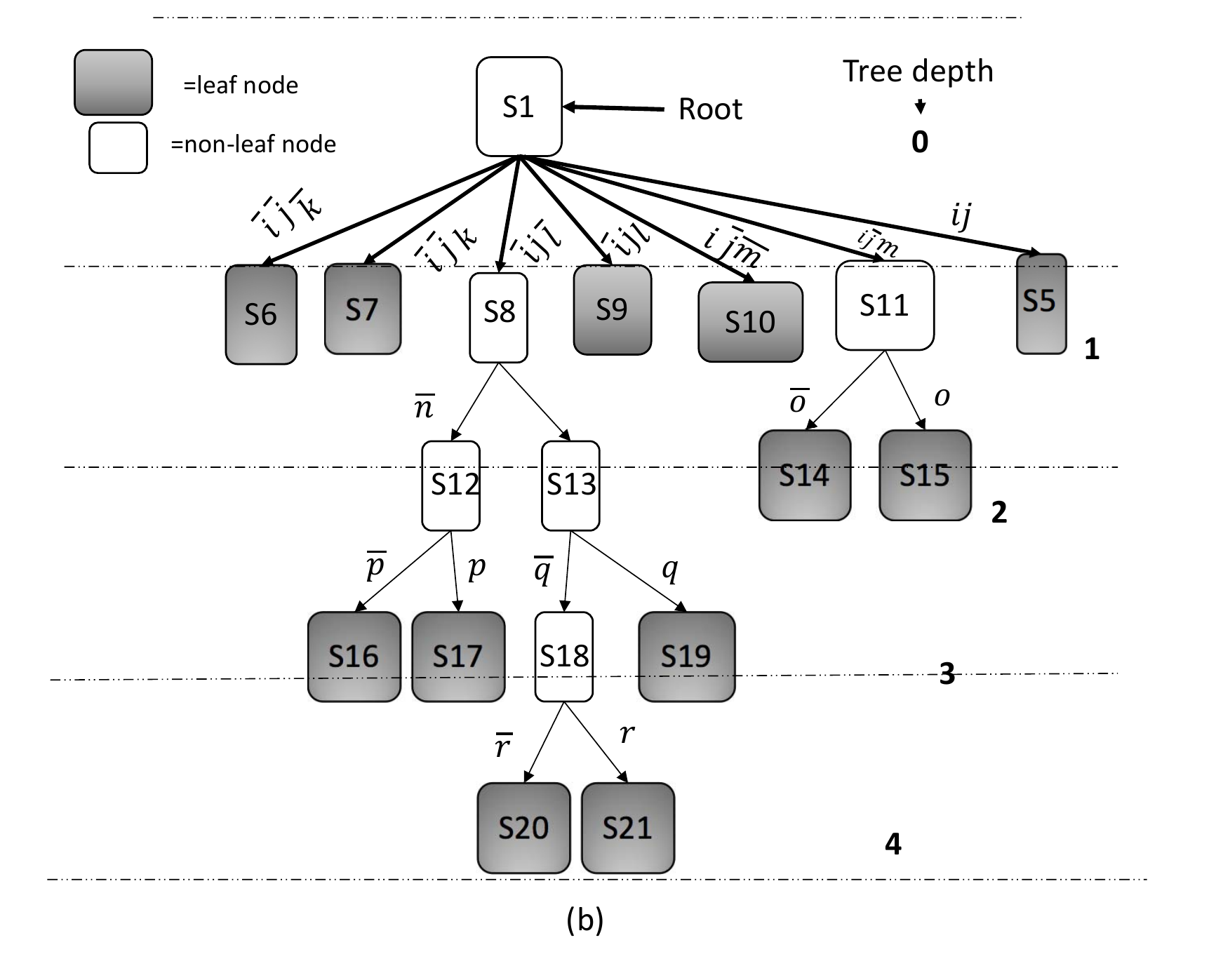}
 	\caption{Multibit tries derivation from decision tree in Figure~\ref{fig:decision-tree}. Note: the don't care is omitted in the illustrating tree. For example, the edge from State $S1$ to $S5$ should be $ijxxx$.}
 	\label{fig:multibit-trie}\vspace{-3mm}
 \end{figure}

\begin{figure}[h]
 	\centering
 	\includegraphics[width=.95\linewidth]{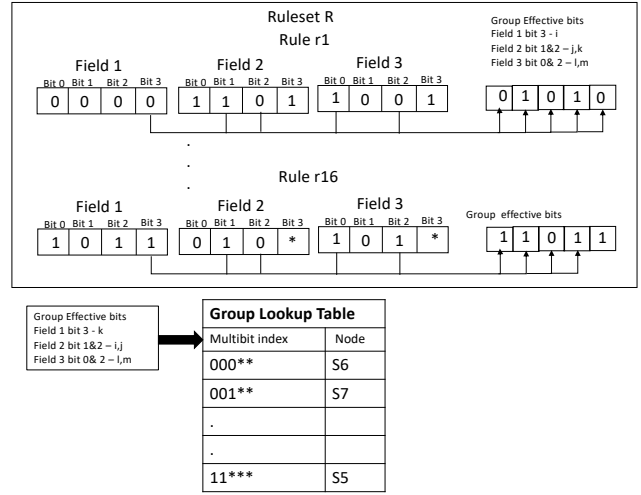}
 	\caption{Methodology of lookup table construction for multibit trie.}
 	\label{fig:lookuptable}\vspace{-3mm}
 \end{figure}

After building the memory, time or both optimized decision tree, we could further reduce the memory and classification time by introducing multibit tries scheme. In order to give a clear explanation of this scheme, a ruleset as shown in Table~\ref{tab:exmaple1} with 3 field which is extracted from 5-field ruleset could be used. The multibit tries scheme is illustrated in Figure~\ref{fig:decision-tree} and Figure~\ref{fig:multibit-trie} and subsequent lookup table construction is shown in Figure~\ref{fig:lookuptable}.


\begin{table}[h]
	\begin{center}
		\small
		\begin{tabular} {|c|c|c|c|}
			\hline
			Rules        & Field 1 & Field 2& Field 3\\ \hline
			$r_1$        & 0010    & 1101    & 1001    \\ \hline
			$r_2$        & 1001    & 000*    & 100*    \\ \hline
			$....$         &   .....     & ......     &  ......     \\ \hline
			$r_{16}$       & 1011    & 010*    & 101*    \\ \hline
		\end{tabular}
		\caption{An example ruleset of 16 rules with 3 fields} \label{tab:exmaple1}
	\end{center}\vspace{-1mm}
\end{table}

As shown in Figure~\ref{fig:decision-tree}, the decision tree consists of a root, terminal or leaf and non-terminal nodes. Algorithm~\ref{alg:TruncatedDecisinTree} describe the actions required to do the proposed multibit tries. It takes decision-tree and  group-binth($th_G$) as input and output a truncated version of the decision tree. All the tree nodes pointers are stored so that they could be traversed in DFS manner(line 1-2). Every non-leaf binth except root is then compared with given $th_G$ for the number of rules attribute (line 4-5). In line 6, if the condition is true, the selected node's pointer from it's parent is reconfigured with selected node's parent's to it's children pointers. After that we delete that node from the tree (line 7).

Following example illustrates this process. $S1$ is the root and $S2, S3, S4, S5$ are spawned by cutting $S1$ in any 5 of the dimensions/ fields or combination of them. For this specific scenario, this cut could be represented by bit $i,j$. In Figure~\ref{fig:lookuptable} it is shown that $i$ is field 2 bit1, and $j$ is field 2 bit2.

 In tree's next level, $S2$, $S3$ and $S4$ are again divided into new node pairs of ($S6,S7$), ($S8,S9$), ($S10,S11$) respectively. This individual cuts could be represented by bit $k$,$l$ \& $m$, where $k$ is field 1 bit 3, $l$ is field 3 bit 0, $m$ is field 3 bit2. Multibit tries scheme enables us to concatenate bit $i,j,k,l $\& $m$ and create a direct relationship between root node $S1$ and $S6,S7,S8,S9,S10,S11$ as shown in Figure~\ref{fig:multibit-trie}. One important thing to mention here is we select to shrink $S8$ and $S11$, but not $S12,13$ or $S18$ because $S8$ and $S11$ contains number of rules more than a given threshold. This threshold is called group-binth. On the other hand, by binth, we represent the threshold number of rules that we used to decide a terminal or leaf node is reached or not. So, introducing group binth concept enables us to build a tree with less memory footprint and lower depth. The reason for lower memory footprint is because we can eliminate intermediate non-terminal nodes in final tree. This also enables to lessen the tree depth which essentially enables a smaller number of memory access to reach of leaf node.


\vspace{-5mm}
\section{Results} \label{sec:results}

\subsection{Experiment Setup} \label{sec:setup}

We used python to build the tree environment, that is the tree and all the actions. We used Proximal policy optimization (PPO)~\cite{SchulmanWDRK17} along with actor-critic algorithm as described in~\cite{Liang-NPC-2019} was used to generate the optimized trees for the rulesets presented in the result section.
%
%
%

\vspace{-3mm}

\subsection{Memory Footprint Result}

\begin{figure}[t]
\centering
\includegraphics[width=.95\linewidth]{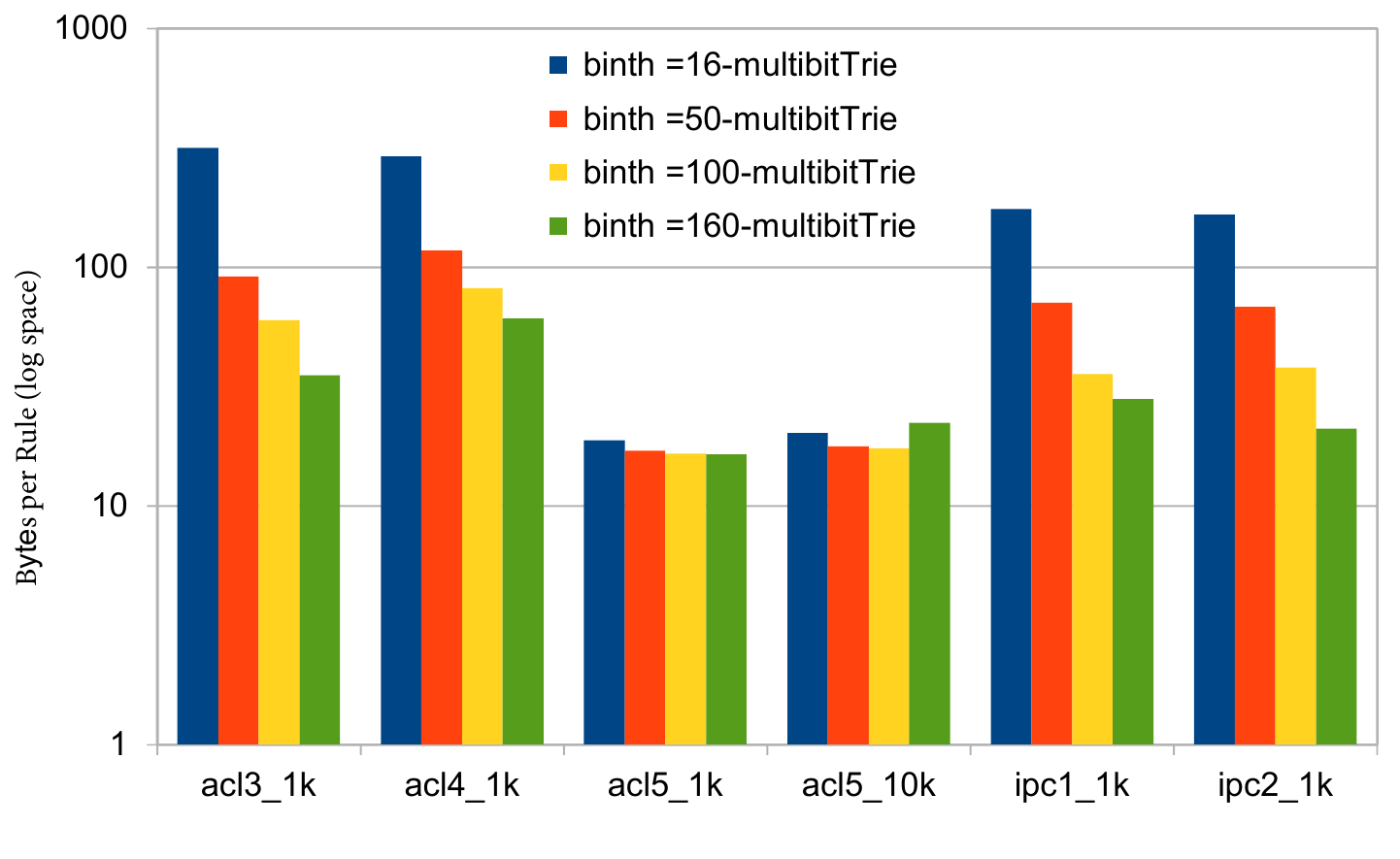}
\vspace{-5mm}\caption{Memory footprint (bytes per rule) for different leaf binth value with group binth equaling to 160.}
\label{fig:memory-binth-ruleset}

\end{figure}

Figure~\ref{fig:memory-binth-ruleset} illustrates the memory footprint for different leaf-binth values for different rulesets. It clearly shows that the higher binth (leaf-binth) value yields a lower memory footprint. This is because higher binth value implies that during construction of a tree, the leaf nodes could be achieved in earlier state (i.e., with fewer cuts) than for a lower binth value, making the total number of nodes in the tree smaller.

\subsection{Performance Result}

\begin{figure}
\centering
\includegraphics[width=.95\linewidth]{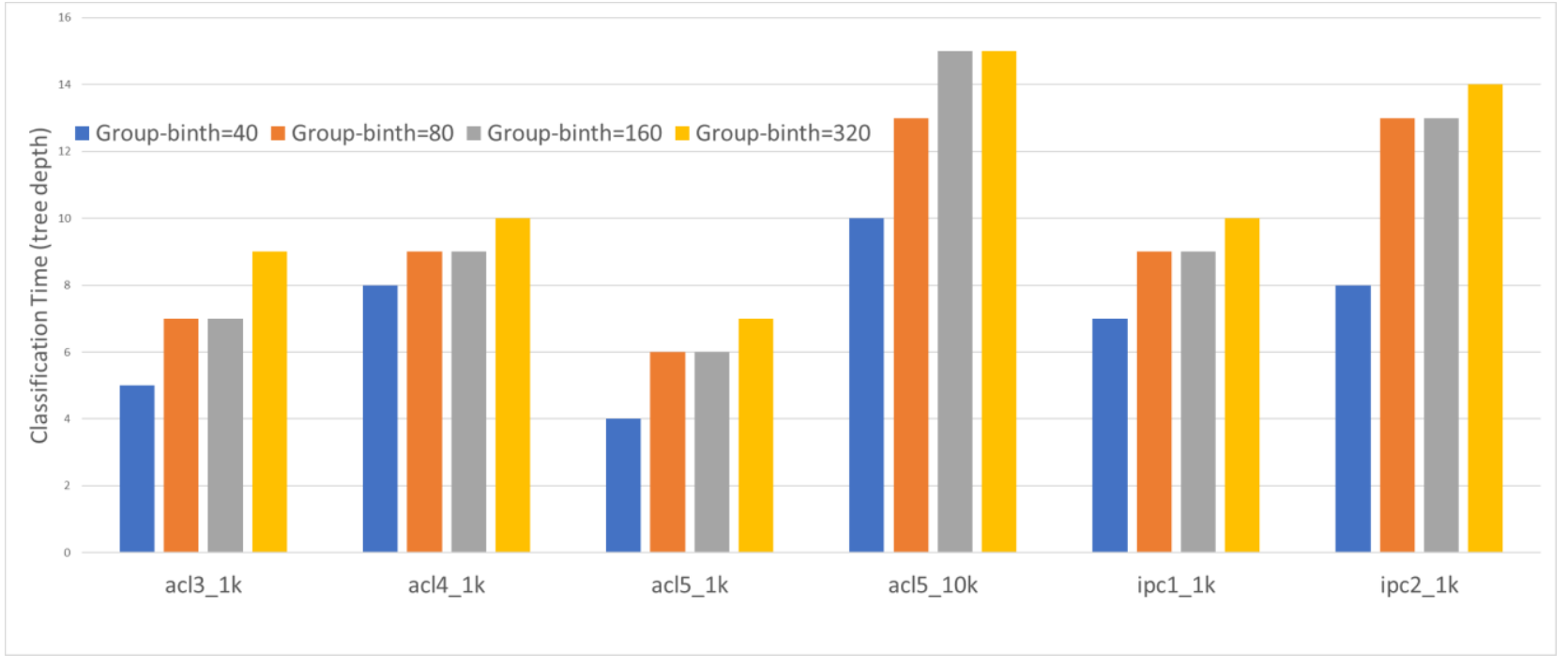}
\vspace{-1mm}\caption{Classification time (tree depth) for varying group binth value. The leaf binth value is 16.}
\label{fig:performance-worst-binth}
\vspace{-3mm}
\end{figure}

\begin{figure}
\centering
\includegraphics[width=.95\linewidth]{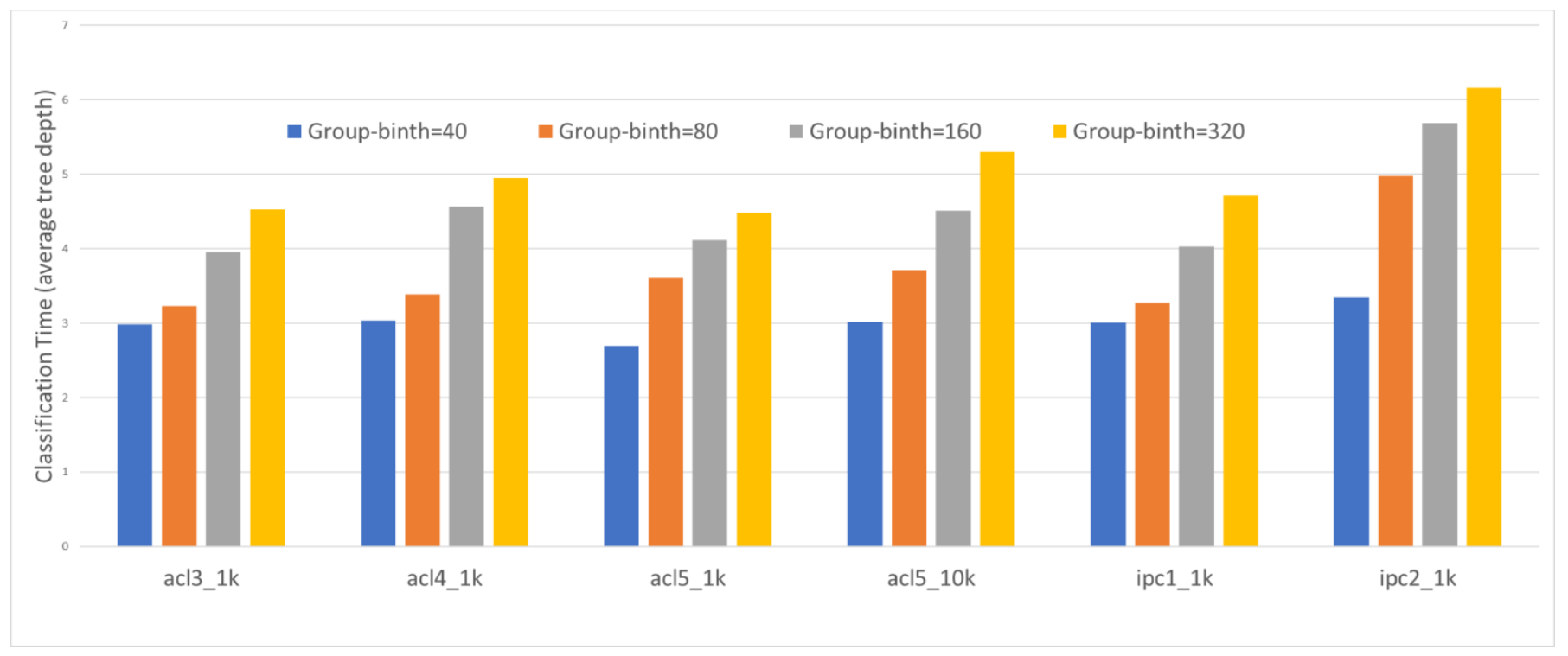}
\vspace{-1mm}\caption{Average classification time (average tree depth) for varying group binth value. The leaf binth value is 16.}
\label{fig:performance-average-binth}
\vspace{-3mm}
\end{figure}

Figure~\ref{fig:performance-worst-binth} describes the effect of varying group-binth in classification time (tree depth) for different rulesets. If we consider to average all the nodes depth (i.e. average tree depth), we will get results described in figure ~\ref{fig:performance-average-binth}. With the higher group-binth value, the less number of non-terminal nodes will be truncated from the decision tree compared with lower-binth. For this reason, with increase in group-binth, both the worst and average case performance decreases (i.e. the generated tree depth for worst case and average tree depth for average case increases).

\begin{table*}[h]
\begin{center}
\caption[]{Performance \& memory requirement for Decision Tree and multibit tries decision tree with leaf-binth value=16 \& group-binth value=40}
\small
\begin{tabular}{|l||c|c||c|c||c|c|}
\hline
&\multicolumn{2}{c||}{memory footprint (bytes per rule)}
&\multicolumn{2}{c||}{\# of worst case memory accesses} &\multicolumn{2}{c|}{\# of average case memory accesses}\\
ruleset &decision tree&multibit decision tree &decision tree &multibit decision tree &decision tree & multibit decision tree  \\
\hline
\hline acl3\_1k  &332.55 &319.91 &9 &5 &4.52 &2.98\\
\hline acl4\_1k  &311.93 &296.88  &11 &8 &6.81 &3.03\\
\hline acl5\_1k  &21.13 &19.75 &9 &4 &5.43 &2.69\\
\hline acl5\_10k &23.204 &21.29 &18 &10 &8.03 &3.01\\
\hline ipc1\_1k  &185.68 &178.03 &10 &7 &5.19 &3.009\\
\hline ipc2\_1k  &182.62 &172.008 &14 &8 &6.15 &3.34\\
\hline
\end{tabular}
\vspace{-3mm}
\label{tab:comparision-results-binth-50}
\end{center}
\end{table*}

\subsection{Comparisons with Decision Tree}

The performance gain and memory footprint with multibit tries scheme is described in Table~\ref{tab:comparision-results-binth-50}. With leaf-binth =16 and group-binth =40, both in worst and average case,the classification time, and memory footprint are improved with our multibit tries scheme.
From the table~\ref{tab:comparision-results-binth-50}, should note that the memory footprint is not significantly lower in multibit tries scheme. This is due to the fact that in multibit tries scheme, we essentially truncate some selected non-leaf nodes. Compare to the size of a leaf node,
a non-leaf node has significantly lower memory requirement as non-leaf node doesn't have any rules stored in them and only contain it's child nodes pointers and some identification and status variables.
All these enables multibit tries scheme to achieve up to 55\% better performance . This is significant improvement in performance in expense of pre-computing the group lookup table (as shown in Figure~\ref{fig:lookuptable}). 

\vspace{-3mm}

\section{Conclusion} \label{sec:conc}

In this paper, we present a learning-based packet classification algorithm and evaluate its performance for varying ruleset. By exploiting the sparsity of rulesets, our algorithm uses a few effective bits to divide a large ruleset into multiple subsets with low rule replication for a lower memory usage at the offline stage. These effective bits are first selected by deep reinforcement learning and then are concatenated based on group binth. Using these effective bits, our algorithm can filter out the majority of rules which do not need to be full-matched to improve the online system performance.

 Our multibit tries classification engine outperforms lookup time both in worst and average case and memory footprint, compared to traditional decision tree without EBs. The performance gain is due to multibit tries enables the classifier to traverse the decision tree several level at a time. The memory reduction is due to 
multibit tries allows us to truncate some selected non-leaf nodes in the decision tree. Preliminary evaluation of small size of ruleset is one limitation of this work. Nevertheless, we believe that our multibit tries is an important step towards learning based high-performance packet classification solution.

\bibliographystyle{IEEEtran}
\bibliography{IEEEabrv,ref}

\vfill

\end{document}